\documentclass[preprint,showkeys,aps,prb]{revtex4}
\usepackage[dvips]{graphicx}
\begin{document}

\title{Order and Chaos in the One-Dimensional $\phi^4$ Model :\\
$N$-Dependence and the Second Law of Thermodynamics}

\author{
William Graham Hoover                      \\
Ruby Valley Research Institute             \\
Highway Contract 60, Box 601               \\
Ruby Valley, Nevada 89833                  \\
}
\author{
Kenichiro Aoki \\
Department of Physics, Hiyoshi Campus, Keio University\\
Hiyoshi, Yokohama 223, Kanagawa, Japan\\
}
\date{\today}


\vspace{0.1cm}

\begin{abstract}
We revisit the equilibrium one-dimensional $\phi^4$ model from the dynamical systems point of
view.  We find an infinite number of periodic orbits which are computationally stable.
At the same time some of the orbits are found to exhibit positive Lyapunov exponents!  The
periodic orbits confine every particle in a periodic chain to trace out either the same or a
mirror-image trajectory in its two-dimensional phase space.  These  ``computationally stable''
sets of pairs of single-particle orbits are either symmetric or antisymmetric to the very last
computational bit. In such a periodic chain the odd-numbered and even-numbered particles'
coordinates and momenta are either identical or differ only in sign. ``Positive Lyapunov
exponents'' can and do result if an infinitesimal perturbation breaking a perfect two-dimensional
antisymmetry is introduced so that the motion expands into a four-dimensional
phase space.  In that extended space a positive exponent results.

We formulate a standard initial condition for the investigation of the
microcanonical chaotic number dependence of the model. We speculate on the uniqueness of the model's
chaotic sea and on the connection of such collections of deterministic and time-reversible states to
the Second Law of Thermodynamics.
\end{abstract}

\maketitle

\section{Introduction}

The study of an anharmonic heat-conducting lattice-dynamics model, the $\phi^4$ model, from
the standpoint of classical statistical mechanics was explored by Aoki and Kusnezov\cite{b1,b2}
and by Hu, Li, and Zhao\cite{b3} in 2000. The Aoki-Kusnezov work led to particularly clear and
easily reproducible illustrations of the phase-space
dimensionality loss found in nonequilibrium steady states as was discussed and illustrated with
Holian, Hoover, Moran, and Posch in 1987\cite{b4,b5,b6}.  Unlike the harmonic chain, in which
heat travels ballistically at the speed of sound, the one-dimensional $\phi^4$ model exhibits
Fourier heat conductivity with a finite large-system limit. This difference to the harmonic
chain is due to the presence of quartic ``on-site'' ``tethering'' potentials, one for each
particle.  These tethers suppress the amplitude of low-frequency waves. We will see that there
is a relatively wide number-dependent energy range within which the tethers induce a chaotic
dynamics.

The Hamiltonian for the one-dimensional $\phi^4$ model is the sum of the kinetic, tethers, and
nearest-neighbor pair-potential energies :
$$
{\cal H} = K + \Phi_{\rm tethers} + \Phi_{\rm pairs} =
\sum_i^N [ \ (p^2_i/2) + (q_i^4/4) \ ] + \sum_{i<j}^{\rm pairs} (q_i-q_j)^2/2 \ .
$$
Here the $\{ \ q \ \}$ represent the displacements of the particles from their static lattice
rest positions. The $\{ \ p = \dot q\ \}$ are the corresponding momenta. The rest length $d$
of the Hooke's-Law springs is irrelevant in this one-dimensional case where it makes no contribution
to the pair-potential part of the equations of motion :
$$
\ddot q_i + q_i^3 = (i+1)d + q_{i+1} - 2(id + q_i) + (i-1)d + q_{i-1} \equiv
q_{i+1} - 2q_i + q_{i-1} \ .
$$
Free, fixed, and periodic boundary conditions are all possibilities.  We mostly choose the
periodic case in which the first and last particles in the chain are linked by a Hooke's-Law
spring so that the resulting ``loop'' is homogeneous and periodic.

Ever since their 1987\cite{b4,b5} work with Brad Holian and Bill Moran,  Harald Posch and Bill
Hoover sought clearcut evidence that the fractal nature of nonequilibrium phase-space
distributions found for small systems persists in larger ones.  The fractal phase-space
structures can be used to explain the Second Law of Thermodynamics in purely mechanical terms
for both microscopic and macroscopic systems.  The fractals not only show the measure-zero nature
of nonequilibrium steady states.  They also clarify the irreversible nature of the unidirectional
repellor-to-attractor phase-space flow. This Second Law connection to fractal structures
can best be established through studies of the dynamical instabilities described by the Lyapunov
spectrum\cite{b6,b7,b8,b9,b10,b11}.

The Lyapunov spectrum $\{ \ \lambda_i \ \}$ has a number of exponents equal to the dimensionality
of the phase space, for which we use the symbol $D$ . The exponents describe the virtual growth
and decay rates parallel to the orthogonal axes of a comoving and corotating phase-space
hypersphere.  The exponents are ordered according to their long-time-averaged values, beginning
with the largest, $\lambda_1$ and ending up with the smallest $\lambda_D$ . $\lambda_1$
describes the time-averaged rate at which two nearby trajectories tend to separate , $\lambda_1
= \langle \ \dot \delta/\delta \ \rangle $ .  We call these rates ``virtual'' because the
numerical algorithms used to measure them maintain trajectory separations by rescaling or by
using Lagrange-multiplier constraints\cite{b8}. Sums of the first $n$ exponents describe the
growth and decay rates of $n$-dimensional comoving and corotating phase-space balls. In the
equilibrium case of pure Hamiltonian mechanics Liouville's Theorem, $\dot f(t) \equiv 0$, along
with the comoving conservation of the phase-space probability, $f\otimes$, implies that the sum
of all the Lyapunov exponents is precisely zero :
$$
\dot f = 0 \ {\rm and} \ (d/dt)(f\otimes) \equiv 0 \longrightarrow \dot f \otimes + f \dot \otimes =
0 + 0 \longrightarrow \dot \otimes = 0 \ .
$$ 
$$
\langle \ (\dot f/f)_t \ \rangle = \langle \ [ \ (\dot f(t)/f(t) \ ] \ \rangle = - \langle 
\ (\dot \otimes(t)/\otimes)(t) \ \rangle = \langle \ -\sum_i^D \lambda_i(t) \ \rangle =
-\sum_i^D \lambda_i \equiv 0 \ .
$$
Here $\lambda_i(t)$ is the $i$th instantaneous exponent and $\lambda_i$ is its time average.
Hamiltonian long-time-averaged exponents occur in equal and opposite pairs,
 $\{ \ \pm\lambda \ \}$ , corresponding to the time reversibility of the motion equations.
Expansion and contraction exchange places in a reversed Hamiltonian flow.
{\bf Figure 1} shows the 32 Lyapunov exponents for two periodic 16-particle $\phi^4$ chains. In
both cases the two vanishing exponents correspond to the lack of growth or decay in the direction
of the phase-space trajectory and in the direction perpendicular to the 32-dimensional energy
surface $E = {\cal H}$ .\\
\noindent

\begin{figure}
\includegraphics[width=4.5in,angle=+90.]{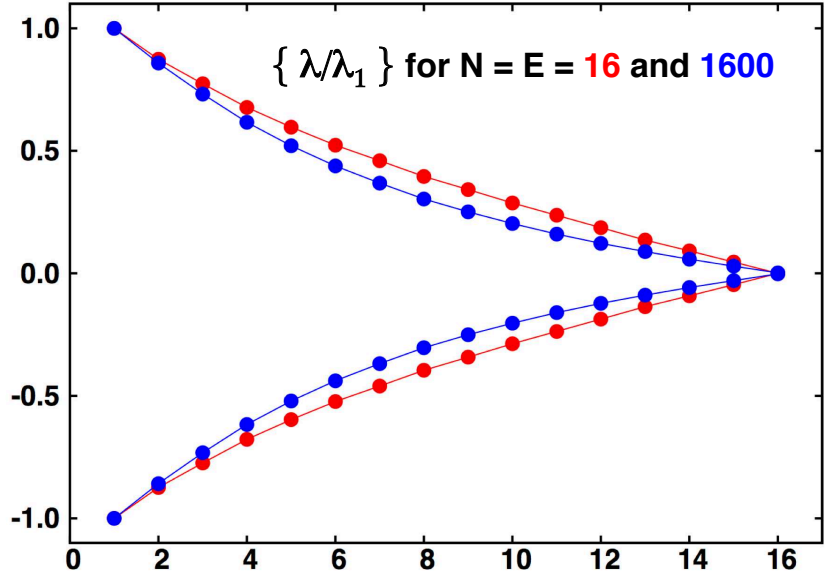}
\caption{
The 16 pairs of Lyapunov exponents for chaotic and periodic 16-body $\phi^4$ chains, ``loops'',
with $(E/N) = 1 \ {\rm and} \ 100$ .  The spectra have been divided by the largest Lyapunov
exponents, $\lambda_1 = 0.0746 \ {\rm and} \ 0.242$ , respectively. The red/blue points
correspond to 16/1600 respectively.\\
}
\end{figure}

The nonequilibrium case is quite different\cite{b4,b5,b6,b9,b10,b11}. It does seems likely that
this $\phi^4$ model will prove useful for future nonequilibrium studies involving the
thermodynamics of heat transfer.  Accordingly we review our current knowledge of nonequilibrium
aspects of the model here.  Velocity gradients or thermal gradients induced or maintained
by deterministic thermostats invariably lead to a breaking of time symmetry.  Away from
equilibrium  the thermostated time-averaged rate of change of the phase volume
$\langle \ \dot \otimes \ \rangle$ is invariably negative.  The thermostated phase volume
shrinks onto a stationary strange attractor.  The attractor has a {\it fractional} ``information
dimension'' $D_I$ less than that of the phase volume, $D$ . The information dimension involves
a sum over phase-space bins of linear size $\epsilon$ :
$$
D_I = \sum_{\rm bins} [ \ P \ln (P)/\ln (\epsilon) \ ] \
( \ {\rm in \ the \ small-}\epsilon \ {\rm limit} \ ) \ .
$$
The bin probabilities are normalized : $\sum P \equiv 1$ .

A direct measurement of the information dimension is impractical for problems with more than
three or four phase-space dimensions because the number of bins becomes prohibitive.
Accordingly Kaplan and Yorke suggested a handy approximation $D_{KY}$ to the information
dimension :
The Kaplan-Yorke approximation is determined by linear interpolation between the dimensionality
of the highest-dimensional expanding ball $(D_e)$ and the dimensionality of the
lowest-dimensional contracting ball $D_c = (D_e+1)$ :
$$
\sum_i^{D_e}\lambda_i > 0 > \sum_i^{{D_e}+1}\lambda_i = \sum_i^{D_c}\lambda_i
\longleftrightarrow D_e < D_{KY} \simeq D_I <  D_c = D_e+1 \ .
$$

When $\lambda_1 > 0$ and the ``Kaplan-Yorke'' fractal dimension $D_{KY}$ of the distribution
is less than that of the phase space the distribution of trajectory points occupies a ``strange
attractor''.  In such cases, the probability of finding states violating the Second Law of
Thermodynamics {\it vanishes} rather than just being small\cite{b4}.  The ``volumes'' of
fractals are {\it zero} in their embedding spaces.

Aoki and Kusnezov's $\phi^4$ model provides many far-from-equilibrium examples of the relatively
large dimensionality loss $\Delta D = D - D_I \simeq D - D_{KY}$ .  For example two-dimensional
square-lattice $\phi^4$ models with 64, 100, and 144 particles, with one corner hot and
another, diagonally opposite, cold, gave dimensionality losses $\Delta D$ of 12.5\cite{b6},
21.6\cite{b10}, and 33.8\cite{b10}.  In their recent book the Hoovers extended the one-dimensional
calculations to 24- and 32-particle chains with dimensionality losses of $\Delta D \simeq 35$
out of 48+2 and $\Delta D \simeq 43$ out of 64+2 phase-space dimensions\cite{b11}.

In the present work we characterize the Lyapunov instability of equilibrium loops and chains
from the standpoint of dynamical systems theory, seeking to outline the region in which chaos
is present and to explore its characteristics. In Section II we consider a standard initial
state and discuss tests for chaos based on the largest Lyapunov exponent and the distribution
of kinetic temperature $\{ \ p^2 \ \}$ . Detailed results are given in Section III. Our
conclusions and recommendations for further work are summarized in Section IV.

\section{A Convenient Initial Condition for Chaotic Chains}

The restlength of the nearest-neighbor springs is irrelevant in one dimension.  Without loss
of generality it can be chosen equal to zero with the $\{ \ q \ \}$ representing displacements
about a common origin.  Evidently $\phi^4$ thermodynamics depends upon only one intensive
variable, the internal energy $(E/N)$ [ or, equally well, the kinetic temperature,
$\langle \ p^2 \ \rangle$, or the specific potential energy, $(\Phi/N)$ ] but {\it not} at all upon a specific volume ( length ) or density
variable.  To choose an initial condition consistent with a particular
conserved energy $E$ it is simplest to follow a two-step process.  First, choose all of the $N$
momenta randomly, using the random number generator described below. The sign of the momenta
is unimportant as momentum is not conserved by the $\phi^4$ model. Next, rescale the momenta
so as to generate the desired initial energy $E$ . Initially, but not for long, the total
energy is all kinetic : $E = K_{t=0} = \sum (p^2/2)_{t=0}$. For convenience in our numerical
work we choose the mass and Boltzmann's constant both equal to unity and integrate the
equations of motion with a fourth-order Runge-Kutta integrator, choosing the timestep such
that the rms single-step integration error is of order $10^{-10}$ .  In doubtful or surprising
cases an adaptive integrator comparing the integration over a timestep $dt$ to two successive
integrations with timesteps $(dt/2)$ is useful.

An alternative to Hamiltonian mechanics is ``thermostated'' mechanics which by now has a huge
30-years' literature.  We choose to use the simplest possible ( Nos\'e-Hoover ) thermostat(s).
To thermostat an $N$-body periodic $\phi^4$ loop it is only necessary to thermostat one of the $N$
particles at the desired temperature $T$.  In nonequilibrium simulations it is usual to thermostat
two particles, one ``hot'' and one ``cold'', at the two ends of an $N$-body chain. The equations
of motion for any thermostated particle, either at equilibrium or away, include an extra thermostat
force imposed by a friction coefficient or ``control variable'' $\zeta$ :
$$
F_{\rm NH} = - (\zeta p)_{\rm NH} \ ; \
\dot \zeta_{\rm NH} = p_{\rm NH}^2 - T_{\rm NH} \
[ \ {\rm Nos\acute{e}-Hoover \ Thermostat} \ ] \ .
$$
We will apply this thermostat to our equilibrium simulations in Section IIID.

\subsection{Definition of Kinetic Temperature Through the Ideal-Gas Thermometer}
The definition of ``kinetic temperature'' $\langle \ p^2 \ \rangle \equiv T$ and our exclusive
use of that temperature in this work, is based on the thermodynamic definition of temperature
in terms of an ideal-gas thermometer.  Conceptually such a thermometer is made up of many tiny
fast-moving particles.  Frequent collisions ensure that the thermometer has always a
Maxwell-Boltzmann distribution of momenta, $f(p) \propto e^{-p^2/2T}$ . It is a straightforward
kinetic-theory exercise to show that a massive particle's interaction with such a thermometer
results in a frictional force on the heavy particle, proportional to its velocity.  Further
a similar calculation for our one-dimensional case shows that a heavy particle loses energy
to an ideal-gas thermometer if its mean squared velocity exceeds $(kT/M)$ where $T$ is the
ideal-gas temperature and $M$ is the massive particle's mass\cite{b13}.  Likewise the heavy
particle gains energy if $(kT/M)$ exceeds its mean squared velocity.  Defining the temperature
of a particle as that of the thermometer which neither gains nor loses energy due to collisions
provides an unambiguous mechanical definition of that particle's temperature.  This definition
is fully consistent with equilibrium thermodynamics and also facilitates the analysis of
nonequilibrium situations involving one or more heat reservoirs. Such reservoirs are simply
large versions of the ideal-gas thermometer.

\subsection{Definition and Computation of the Largest Lyapunov Exponent}

In any case, at a fixed energy $E$, or thermostated at one equilibrium temperature $T$, or
at two nonequilibrium temperatures $T_{\rm hot \ and \ cold}$, there are at
least four distinct ways to determine the largest Lyapunov exponent.  From the conceptual 
standpoint all four involve following the motion of two similar systems, the ``reference''
trajectory which is unperturbed, and a nearby ``satellite'', which is constrained to evolve
at a fixed separation from the reference. The satellite trajectory can be described in phase
space ( by solving identical equations of motion ) or in ``tangent space'' where the offset
is infinitesimal and the satellite equations of motion are linearized with respect to the
offset, $\{ \ \delta q,\delta p,\delta \zeta \ \}$ .  The constant-offset constraint
can be imposed by rescaling at the end of every timestep or by including an extra Lagrange
multiplier\cite{b7,b8} in the satellite motion equations.  For finite separation a convenient
choice is
$$
\delta = 0.000001 =
\sqrt{\sum [ \ (q_s - q_r)^2 + (p_s - p_r)^2 \ ] + (\zeta_s - \zeta_r)^2} \ .
$$
We have used both phase-space and tangent-space methods, both fixed timestep and
variable-timestep Runge-Kutta integrators, compiled from both FORTRAN and C in order to
check our work.  For more details of the Lyapunov algorithms and several examples see Chapter
5 of Reference 11 or the many papers on this subject in the Los Alamos ar$\chi$iv.

\subsection{Random Number Generator}

In many of our simulations we have used the six-line random number generator
{\tt rund(intx,inty)} with the two seeds {\tt intx} and {\tt inty} initially set equal to
zero.  This generator is time-reversible\cite{b12}.  Its forward version is as follows :\\

\noindent
{\tt
      i = 1029*intx + 1731\\
      j = i + 1029*inty + 507*intx - 1731\\
      intx = mod(i,2048)\\
      j = j + (i - intx)/2048\\
      inty = mod(j,2048)\\
      rund = (intx + 2048*inty)/4194304.0\\
}

\noindent
$2^{22} = 4 194 304$ pseudorandom numbers are generated before the algorithm repeats.

We recommend the use of this six-line generator for three reasons : [ 1 ] it is simple to
implement ; [ 2 ] it is reproducible, so that colleagues working with different hardware
or software can readily replicate each others' work ; [ 3 ] it is ``time-reversible'' so
that the seed-dependent sequence of $2^{22}$ pseudorandom numbers can be extended either
forward or backward in ``time''.  This last property of time reversibility was established
by Federico Ricci-Tersenghi in his solution of the 2013 Ian Snook Prize Problem\cite{b12}.
This property makes it possible to follow ``stochastic'' evolutions of few-body or
many-body dynamics backward in time .

At sufficiently low temperatures where the quartic potential can play no role  the $\phi^4$
model motion becomes harmonic.  In this case the lowest
frequency corresponds to a wavelength of $N$ for periodic boundary conditions and $2N+2$
for fixed boundaries, with just $N$ motion equations for the coordinates and for the momenta.
We have also used free boundaries at the endpoints which likewise have $N$ equations each
for the coordinates and momenta.  The amplitude of the harmonic motion follows from the
harmonic oscillator relation for a vibrational normal mode of frequency $\omega$ with the
energy equally divided among the system's $N$ modes : 
$$
\sqrt{\langle \ q^2 \ \rangle = (kT/m\omega^2)} \simeq \sqrt{(E/N)(2N)^2} \simeq \sqrt{(EN)}
\simeq \sqrt{TN^2} \ .
$$
At temperatures $T$ higher than $(1/N^2)$ the long wavelength harmonic waves are scattered to
higher frequencies by the tethering potentials.

\subsection{Monte-Carlo Determination of the Chaotic Measure}

Over most of the energy range chains or loops of length 8 or more are typically chaotic, but
this cannot be the case at very low or very high energies.  To determine the relative
measures of the tori and the chaotic sea we have used the following idea :\\
\noindent
[ 1 ] Use scaled random numbers from the generator in the previous section to start a
simulation with a desired energy, initially wholly kinetic.\\
\noindent
[ 2 ] Measure the Lyapunov exponent for 2 000 000 000 timesteps.\\
\noindent
[ 3 ] Make $(N/2)$ vectors of length $r = | \ (p_i - p_{i+1}) \ |$ with the $(N/2)$ distinct
pairs ( where $i$ is odd ) of
adjacent momenta, rotate each vector through a random angle $\theta$ between 0 and $(\pi/2)$ .
Setting the momenta equal to $[ \ r\sin(\theta),r\cos(\theta) \ ]$ provides a new initial
condition with the same energy as before.\\
\noindent
[ 4 ] Repeat steps 2 and 3 above for a sufficient number of trials ( 40 is reasonable ).\\

Because this procedure satisfies ergodicity ( any isoenergetic configuration is able to
be accessed ) and ``detailed balance'' ( the probability of
going from state $I$ to state $J$ is the same as that from $J$ to $I$ ) because the 
algorithm is time reversible.  Thus its implementation will ( ``eventually'' ) converge
to Gibbs' microcanonical ( constant-energy ) average.  Let us turn to an exploration of
results obtained with the methods just described.  Our main goal is to determine the
extent of the chaos in the $\phi^4$ model.  In the course of that work we encountered
several surprises.  They are included in what follows.

\section{Numerical Results}

{\bf Figure 2} shows the dependence of the largest Lyapunov exponent on $(E/N)$ for $N=16$
and $N=500$ .  These systems are sufficiently large and energetic that our standard initial
condition leads to chaos over a wide range of energies.  It is remarkable that the simple
$\phi^4$ model has a readily-accessible chaotic range of about ten orders of magnitude in
the energy.\\

\begin{figure}
\includegraphics[width=4.5in,angle=+90.]{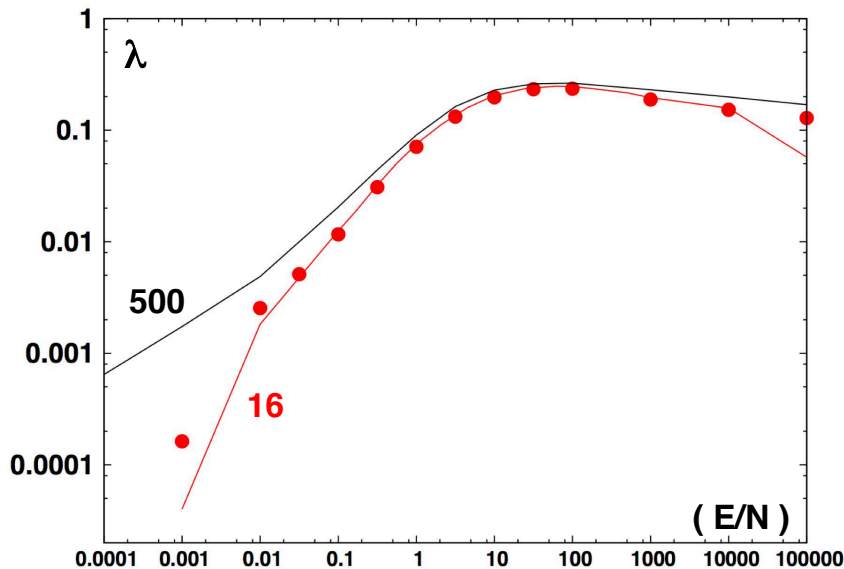}
\caption{
The energy dependence of the largest Lyapunov exponent for periodic systems of 16 and 500
particles are shown as lines.  Data using fixed boundary conditions with 16 moving particles
and two fixed boundary particles are shown as filled circles. All these simulations were
initiated with vanishing coordinates $\{ \ q \ \}$ and with randomly chosen initial
velocities scaled to provide the desired energy.  The trajectories were integrated for
sufficient time that the uncertainties in the $\{ \ \lambda_1 \ \}$ are smaller than the size
of the filled circles.\\
}
\end{figure}

\subsection{The Equilibrium Thermal Equation of State}
At very low temperatures the motion is harmonic so that the energy approaches the
equipartition result, $(E/2) = K = \Phi = NT/2$ , where $K$ and $\Phi$ are the kinetic
and potential energies.  In the opposite high-temperature limit,
$$
\langle \ (q^4/4) \ \rangle \simeq T \rightarrow \Phi \simeq (NT/4)  \ .
$$
For orientation notice that {\bf Figure 3} shows that the kinetic and potential
energies satisfy equipartition ( they are equal ) at low temperature.  At high temperature
where the configurational integral $\int e^{-\Phi/kT}dq \simeq T^{1/4}$ the slope,
$d\Phi/dT$ changes from (1/2) to (1/4) .  For the plot we have used states from the chaotic
sea.  From the {\it rigor mortis} standpoint there are also an infinite number of zero-measure
periodic orbits, mostly unstable. Some of them are stable, surrounded by small-measure families
of tori.  We will encounter both the unstable and the stable cases in studying the smallest
interesting case, $N=2$ .\\
\noindent

\begin{figure}
\includegraphics[width=4.5in,angle=+90.]{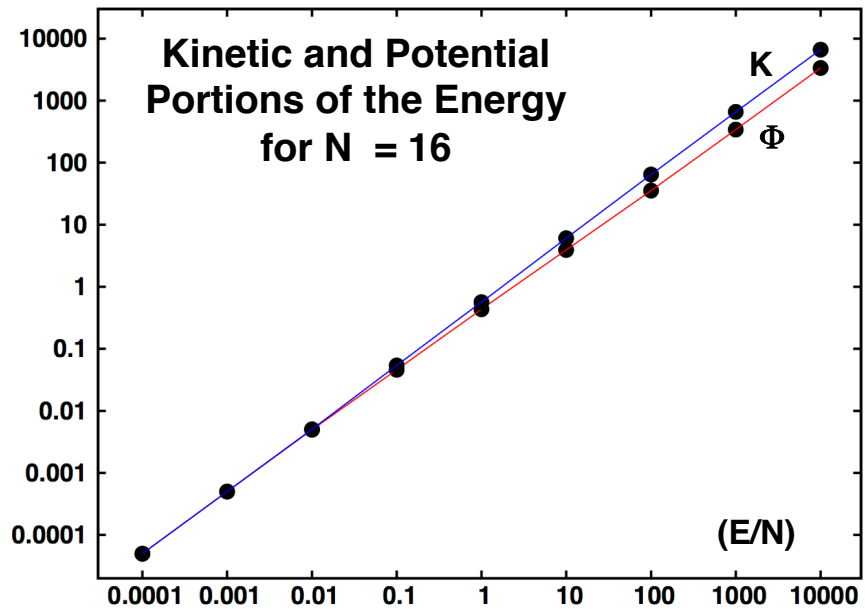}
\caption{
The upper curve shows the variation of kinetic energy per particle and the lower curve
the variation of potential energy per particle with the abscissa values of the total
energy per particle.  The low-temperature equipartition and the high-temperature ratio
of energies correspond to harmonic motion and quartic-potential oscillation respectively.
The data were taken from periodic simulations with $N=16$ .\\
}
\end{figure}

\subsection{$N=2$, the Minimal Case for Chaos}

We begin with the smallest system for which chaos is possible, a pair of one-dimensional
particles. We choose to examine the periodic case, imagining that there are two parallel
Hooke's-Law springs joining the pair :
$$              
{\cal H} = (1/2)(p_1^2 + p_2^2) + (q_1-q_2)^2 + (1/4)(q_1^4 + q_2^4) \ . 
$$
With the energy fixed by the Hamiltonian motion equations this four-dimensional problem
has the minimum dimensionality for chaos, three.  ``Obviously'' solutions with either
of the two symmetry choices $(q_1,p_1) = \pm(q_2,p_2)$ are ``too simple for chaos''.
To see this consider first the symmetric case and set $(q,p) = (q_1,p_1) = (q_2,p_2) $ .
The motion equations are the same for the two particles :
$$
 \dot q = p \ ; \ \dot p = -q^3 \ .
$$
This same result holds for a periodic chain made up of any even number of particles.
In this ``symmetric'' case, with all the particles tracing out the same $(q,p)$ motion the
nearest-neighbor Hooke's-Law potential is constant with its minimum value of zero.  Only
the onsite quartic potential is nonzero.  These are the motion equations in a simple
attractive  quartic potential.

The ``antisymmetric'' case, corresponding to mirror boundary conditions, looks similar. For
two particles or any other even number, all particles obey the same motion equations :
$$
 \dot q = p \ ; \ \dot p = -4q - q^3 \ .
$$
This antisymmetric case describes periodic oscillations in an attractive
potential only slightly more complicated than the symmetric case.  $(q,p)$ phase-plane
plots of both periodic orbits are shown in {\bf Figure 4}.  To avoid overlaps the
particle coordinates $q_1$ and $q_2$ have been shifted to the left and right by 3 .\\
\noindent

\begin{figure}
\includegraphics[width=4.5in,angle=+90.]{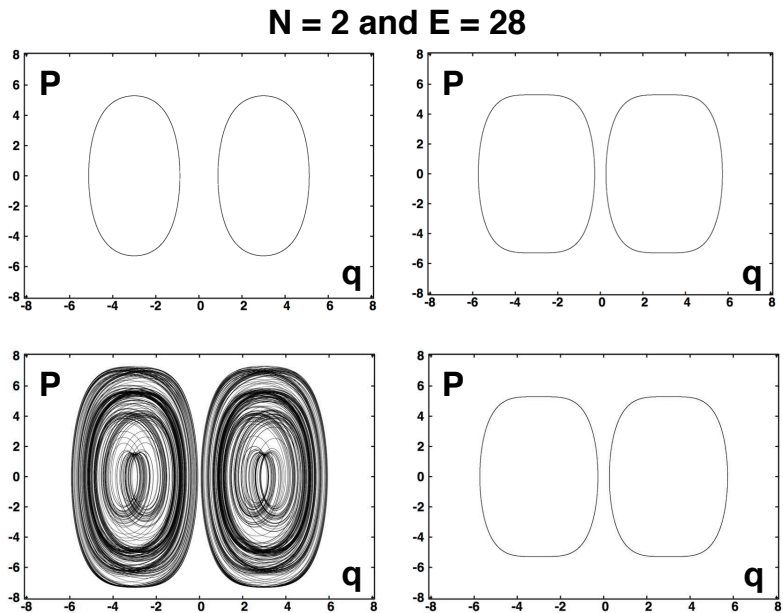}
\caption{
The antisymmetric ( on the left ) and symmetric ( on the right ) phase-plane orbits are
In the symmetric case, with all particles tracing out the same $(q,p)$ motion the
nearest-neighbor Hooke's-Law potential is constant with its minimum value of zero.  Only
the onsite quartic potential is nonzero, as is shown here for an energy of 28.  Both
of the top-row orbits, as well as their periodic repetitions, are computationally stable
to the very last bit.  In contrast, adding a small perturbation to any of the four variables
opens up a four-dimensional phase space and reveals that the antisymmetric case is then
Lyapunov unstable ( as shown below at the left ).  The symmetric case remains stable,
revealing the existence of a torus with nonzero measure in that symmetric case.\\
}
\end{figure}

From the mathematical standpoint the symmetric and the antisymmetric problems are both
equivalent to one-body problems tracing out periodic orbits in a two-dimensional $(q,p)$
phase space and as such are immune to chaos.  But this brief discussion ruling out chaos
in two dimensions is completely erroneous in four!  After all it seems possible that the
symmetric and antisymmetric orbits in the original four-dimensional phase space could
themselves be unstable to small perturbations which are inaccessible in the simpler
two-dimensional symmetrized spaces. In such a case double-precision roundoff errors
{\it might} be enough to provide a seed for instability on the three-dimensional ( as
opposed to one-dimensional ) energy surface. Numerical exploration shows that an energy
$E=15$ is enough for chaos with a positive $\lambda_1$ in the full four-dimensional
$(q_1,p_1,q_2,p_2)$ space.

If we start out with the antisymmetric initial condition of {\bf  Figure 4} we
find rapid convergence of the largest Lyapunov exponent to a value of order unity :
$$                                                                                                    
\{ \ q_1,p_1,q_2,p_2 \ \} \simeq \{ \ +2,+2,-2,-2 \ \} \longrightarrow \lambda_1 = 0.617 \ .
$$ 
Apart from a phase shift we expect this initial condition to correspond equally well to
the purely-kinetic initial condition. Computation shows that this is true :
$$
\{ \ q_1,p_1,q_2,p_2 \ \} \simeq \{ \ 0, +\sqrt{28},0, -\sqrt{28} \ \}
\longrightarrow \lambda_1 = 0.617 \ .
$$
The Runge-Kutta integrator, as interpreted by FORTRAN or C is certainly not
perfect in a mathematical sense.  It isn't even time-reversible.  But it does preserve
{\it symmetry} very nicely ( even perfectly ) as a consequence of arithmetic operations where
only the sign of the numbers is changed.  This symmetry can be lost if the rest lengths of
the springs are incorporated in the equations of motion. {\it Displacement} coordinates
$\{ \ q \ \}$ are advantageous !\\
\noindent

\begin{figure}
\includegraphics[width=4.5in,angle=+90.]{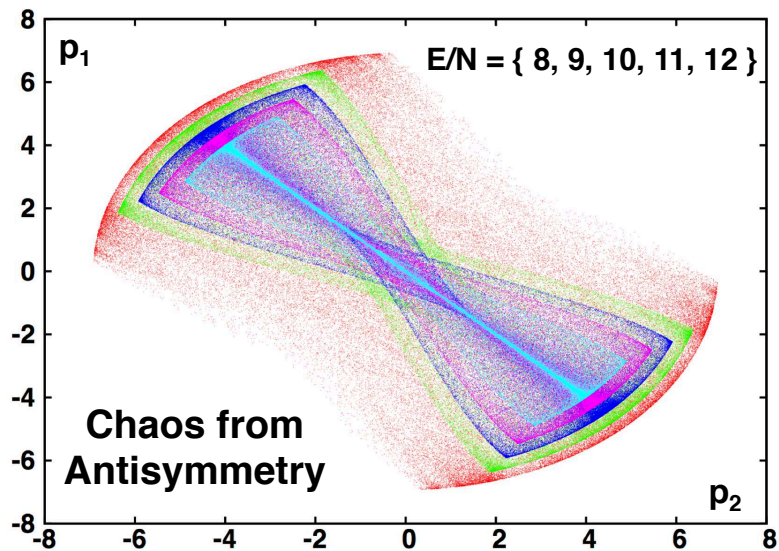}
\caption{
Antisymmetric chaos broadens the correlation of the momenta $p_1+p_2 = 0$ as the energy is
increased.  The two momenta are plotted for energies of 16, 18, 20, 22, and 24.  The transition
from order to chaos occurs near $E=15$ .\\
}
\end{figure}

{\bf Figure 5} illustrates the growth of chaos in the unstable antisymmetric case.  At
low energy the momenta $p_1$ and $p_2$ sum to zero in a regular periodic motion.  Increasing
energy eventually breaks the perfect correlation and gives rise to the increasing chaos seen
in the Figure.  For energies less than fifteen, so that $(E/N) < 7.5$ , the $p_2(p_1)$
correlation is perfect, corresponding to the straight line $p_1 + p_2 = 0$ .

\subsection{Anomalous Orbits for More Pairs of Particles}

It is easy to verify that simulations repeating the same starting condition as above,
$$
\{ \ q,p,q,p \ \} =  \{ \ +2,+2,-2,-2,+2,+2,-2,-2,+2,+2,-2,-2, \ \dots \ \ \} \ ,
$$
where $N = 2, \ 4, \ 6, \ \dots$  all give exactly the same $(q,p)$ plots for every
particle and all give exactly the same Lyapunov exponent, $\lambda_1 = 0.617$ for $(E/N) = 14$ .
See {\bf Figure 6} for $(E/N) = 1,000,000$ .\\

\begin{figure}
\includegraphics[width=4.5in,angle=+90.]{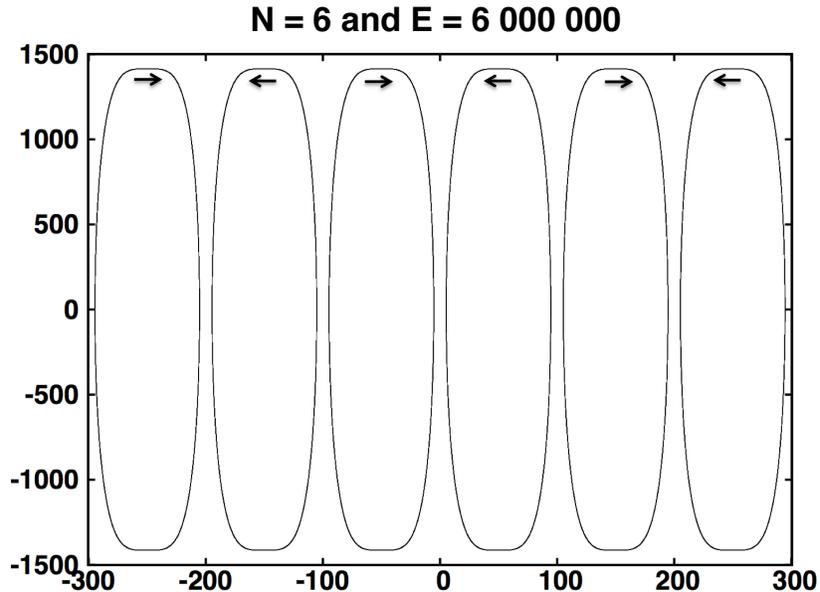}
\caption{
The antisymmetric $(q,p)$ solution applies to any even number of $\phi^4$ particles.  Here
we show the ( offset to avoid overlaps ) phase-plane plots for a periodic system of
six particles.  We emphasize the perfect computational stability, to the very last bit,
of such an orbit.  The case illustrated has $(E/N) = 1,000,000$ .  The corresponding
maximum Lyapunov exponent is 1.153 .\\
}
\end{figure}

What is a bit surprising is that a small perturbation, say $10^{-15}$ , totally changes
things. A nonzero perturbation out of the $(q,p)$ plane can break the antisymmetry.  Such
a perturbation provides a Lyapunov exponent that is not particularly
stable and is considerably smaller, on the order of 0.1 , than the exponent on the
unperturbed periodic orbit.  Evidently the precisely antisymmetric simulations,
without perturbations, differ only in the signs of the $(q,p)$ not the magnitudes.  Thus
standard double-precision arithmetic can maintain perfect antisymmetry and periodicity
with no hint of chaos.  On the other hand the nearby ( perturbed ) satellite trajectory
senses a Lyapunov exponent of 0.617 .  That exponent has nothing to do with a chaotic-sea
average.  It is instead simply the mean value of $\lambda_1(t)$ adjacent to the
underlying periodic orbit.  The symmetric case is less interesting.  Even with an energy
of $10^6$ the symmetric Lyapunov exponent is only 0.03.  With an energy of $10^5$ the
exponent is negligibly small, most likely zero.

For comparison we include another initial condition, neither symmetric nor antisymmetric,
but still with the same initial energy $(E/N) = 14$ :
$$
\{ \ q_1,p_1,q_2,p_2 \ \} \simeq \{ \ +2,+2,-2,+2 \ \} \longrightarrow \lambda_1 = 0.08_6 \ ,
$$
This initial condition evidently samples the chaotic sea rather than just the neighborhood
of a periodic orbit ( we avoid calling the periodic orbits ``stable'' or ``unstable'' as
this is not useful terminology in the two-body case ). An antisymmetric initial condition
with a smaller perturbation should ( we think ) sample the same chaotic sea.  The result
of a computation with a billion timesteps of 0.001 each is $\lambda_1 = 0.08_8$, justifying
our expectation.  In summary the two-particle case ( and the $2N$-particle cases ) exhibit
something interesting, a periodic orbit periodic to machine precision, stable computationally
for so long as the electricity flows, but in the neighborhood of a highly-unstable portion
of the chaotic sea.  

A little reflection suggests that there may well be families of periodic orbits related
to all the normal modes of a chain.  The next step up from $N=2$ is $N=3$ , which exhibits
a computationally perfect symmetry of the type
$$
( \ 0, \ +2, \ -2 ) = ( \ p_1, \ p_2, \ p_3) \ {\rm with}
\ ( \ q_1, \ q_2, \ q_3) = ( \ 0,\ 0,\ 0 \ ) \ \longrightarrow {\cal H} = 4 \ .
$$
This robust periodic solution has a Lyapunov exponent of $0.13_6$, the same order of magnitude
as in the similar two-body solution.  Because the first particle is motionless such a solution
satisfies both the periodic and the fixed boundary conditions.  Such stable periodic orbits
with positive Lyapunov exponents are a fertile field for additional research.  Without pursuing
that subject further here we turn now to another more manageable set of interesting problems, loops
with $N=10, \ 20, \ 40, \ 80, \ 160$ and their approach to the large-system limit.

\subsection{Number-Dependence for Longer Chains and Loops}

With longer chains a systematic number dependence of $\lambda_1$ can be seen.  Seeking
simplicity we begin with periodic chains for which the boundary conditions are homogeneous and do
not single out any part of the system.  For unit energy per particle, $(E/N) = 1$ and in the
chaotic sea, we computed the kinetic energy per particle and the maximum Lyapunov exponent
for  $\phi^4$ loops of 10, 20, 40, 80, and 160  particles.  All of these systems exhibit a
kinetic temperature close to an apparent longchain limit of 1.134 .  Simulations were carried
out using two billion timesteps with a fourth-order Runge-Kutta timestep $dt = 0.001$ .  The
per-particle kinetic energies and Lyapunov exponents we found were as follows :
$$                                                                                          
(K/N)     = \{ \ 0.566_2, \ 0.566_3, \ 0.567_0, \ 0.567_0, \ 0.567_0 \ \} \ ; \             
$$
$$                                                                                          
\lambda_1 = \{ \ 0.0666, \ 0.0767, \ 0.0810, \ 0.0843, \ 0.0871 \ \} \ .                       
$$
The Lyapunov exponents vary roughly linearly in the inverse loop size while
the kinetic energy ( or temperature ) has a variation smaller by two orders
of magnitude. The 31\% increase in $\lambda_1$ is huge relative to the tiny
increase in temperature with a sixteenfold increase in system size.\\
\noindent

\begin{figure}
\includegraphics[width=4.5in,angle=+90.]{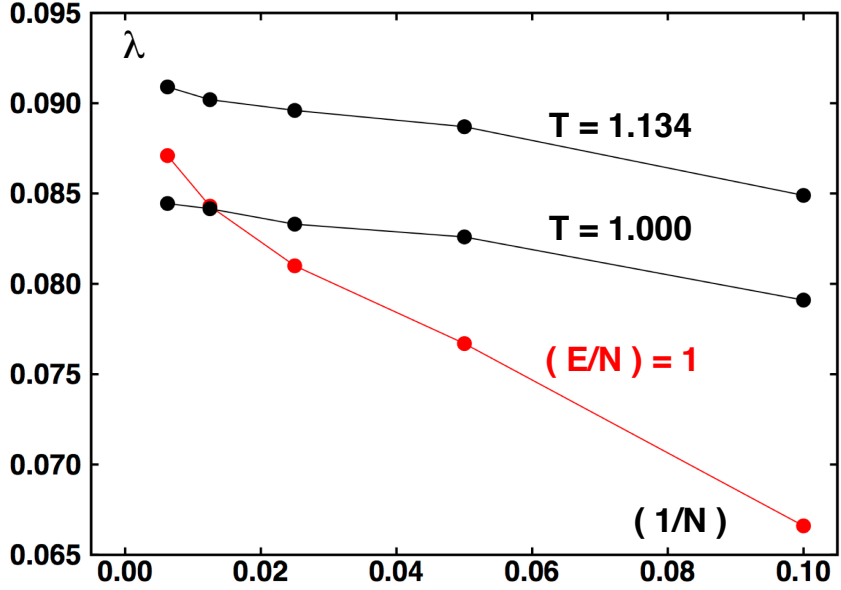}
\caption{
These data indicate that the number dependence of the largest Lyapunov exponent is of order
$(1/N)$ . The steepest curve is for Hamiltonian mechanics with an energy per particle of unity.
The other curves shows $\lambda_1(N)$ for the same system sizes, 10, 20, 40, 80, and 160
particles with a single particle thermostated at a temperature of 1.134 ( at the top ) and
at a temperature of unity ( below ).  The equations of motion for the lone thermostated particle
include the frictional force $-\zeta p$ where $\dot \zeta = p^2 - 1.134$ or $p^2 = 1.000$ .
All these data represent time averages in the chaotic sea with two billion timesteps,
$dt = 0.001$ .\\
}
\end{figure}

To test the sensitivity of the Lyapunov exponent to thermostating we added a
single Nos\'e-Hoover control variable to the motion equations of a single
particle and verified that the chains all came to thermal equilibrium at a kinetic
energy of unity with the motion equation of Particle 1 modified as follows :
$$                                                                                          
\dot p_1 = \dot p_1({\cal H}) - \zeta p_1 \ ; \ \dot \zeta = p^2_1 - 1 \ .                  
$$
The Lyapunov exponents for the thermostated chain at a kinetic temperature of
1 found in this way were :
$$                                                                                          
\lambda_1 = \{ \ 0.0791, \ 0.0826, \ 0.0833, \ 0.0841_6, \ 0.0844_5 \ \}                    
$$ 
At the temperature 1.134 corresponding to unit energy per particle the
largest Lyapunov exponent is somewhat larger, as is shown in {\bf Figure 7} :
$$                                                                                          
\lambda_1 = \{ \ 0.0849, \ 0.0887, \ 0.0896, \ 0.0902, \ 0.0909 \ \}                        
$$
In the present calculations we used $\dot \zeta = p^2 - T$ rather than the
alternative $\dot \zeta = (p^2/T) - 1$ .

\subsection{Dependence of the Chaos on Energy}

{\bf Figure 2} illustrated the dependence of the largest Lyapunov exponent on the
specific energy $(E/N)$.  The falloff at low energy, and eventual disappearance of
the chaotic sea corresponds to the normal-mode structure of the low-energy $\phi^4$
model.  At very low energy, $E \simeq NT < (1/N)$, the initial conditions correspond
to the amplitudes and phases of $N$ normal modes, all of which are periodic in time
so that there is no tendency toward chaos.  At very high energy, where the
Hooke's-Law forces can be ignored relative to the tether forces each particle oscillates
about its lattice site with a regular periodic one-dimensional motion.  For these reasons
the ``interesting'' chaotic range of energies considered here cover nine orders of
magnitude.

The relative measures of the chaotic sea and regular tori vary with system size and
with energy, from (0,1) to (1,0) to (0,1) as the energy varies from zero to order one to infinity.
We have used the Monte Carlo method of Section II.D to determine the
chaotic energy for a sixteen-particle loop, which {\bf Figure 2} showed us is already
close to the large-system limit. At an energy (E/N) = 1,000,000 the Monte Carlo algorithm
returns a chaotic measure of 14/40 .  In the range 0.1 to 1000 all 40 initial conditions
in our microcanonical sample were chaotic. Apart from an early transient ( indicating some
regular measure ) in the Monte Carlo samples with $(E/N) = 0.01$ and 0.001 the measure
there is overwhelmingly chaotic too.

\subsection{Uniqueness and Equilibration of the Chaotic Sea}
 
The realism of the $\phi^4$ model is amazing considering its simplicity.  By considering
hundreds of different initial conditions, randomly chosen but otherwise with equal
energies we have reached the conclusion that the chaotic sea is likely unique.  Given
the number of particles and the energy it appears that there is only one chaotic sea,
not two or three or an infinite number.  Further by considering a more limited number of
chaotic states it appears that their kinetic temperature converges homogeneously :
$$
\langle \ p_1^2 \ \rangle = \langle\ p_2^2\ \rangle = \dots = \langle\ p_N^2\ \rangle \ . 
$$
Without robust thermal equilibrium in the sea we would have to consider the embarassing
possiblity of a violation of the Second Law of Thermodynamics, as is discussed below
among the conclusions and recommentations which have come to us through our studies and
to which we turn next.

\section{Conclusions and Recommendations}

The $\phi^4$ model provides a readily reproducible set of chaotic few-body and
many-body problems where interference from toroidal solutions is minimal.  There
is room for work leading to a quantitative understanding of the first appearance
of chaos at low energies and its last vestige at high.  Preliminary explorations
indicate that the number of nonvanishing exponent pairs varies with energy in the
vicinity of the antisymmetric unstable orbit.

The symmetric and antisymmetric two-body solutions, with the surprising coexistence
of computational stability adjacent to Lyapunov instability was unexpected.  Although
the chaotic sea is a close neighbor to these solutions the identical roundoff errors
for all the even and all the odd numbered particles provides stability adjacent to
chaos.  No doubt other more complex patterns are stabilized by the same
roundoff mechanism, providing nonlinear analogs of the harmonic normal modes.  By
adding dissipative friction to the motion equations the fundamental long wavelength
modes could be captured for any of the choices of boundary conditions, periodic,
fixed, or free.

The mechanical model exhibiting heat flow in response to kinetic temperature gradients
facilitates studies connecting mechanics to thermodynamics.  Because thermodynamics is based
on the ideal gas, with its known Gaussian velocity distribution, entropy, and temperature
links to mechanical systems capable of heat transfer for very small $N$\cite{b14} are
appealing subjects for computational study.

It seems likely to us ( we have so far found no counterexample ) that simulations
in the chaotic sea correspond to global microcanonical thermal equilibria despite their
finite energy and the closeby regular tori with their nonchaotic quasiperiodic time
behavior.  Gibbs' maximum-entropy velocity distribution can be separated from the
highly complex configurational component of the energy surface.  We conjecture that
over a wide range of energies there is a unique chaotic sea in which all  particles
share a common value of the kinetic temperature $\langle \ p^2 \ \rangle$ . If chaotic
solutions were able to provide reproducibly different kinetic temperatures
$\langle \ p^2 \ \rangle$ in different parts of a microcanonical system it would be
possible to violate the Second Law of Thermodynamics by coupling an external Carnot
Cycle to those energy sources and sinks, for the Carnot cycle feeds on the kinetic
energy by exactly the same collisional mechanism as the measurement mechanism of an
 ideal-gas thermometer.

Perhaps the $\langle \ p^2 \ \rangle$ question should be posed differently : ``Under
what conditions will the long-time-averaged kinetic temperatures of all the particles
have the same value ?''
Any robust disparity in the temperatures ( insensitive to small perturbations ) makes
perpetual motion of the second kind possible.  Heat furnished by a ``hotter'' particle
could be used to do work ( with an external Carnot Cycle ) , returning the unused
heat to a ``colder'' one.  Because such a full conversion of heat to work is highly
illegal one would necessarily find that attempts to harness the high-temperature
heat to do work are doomed unless they would simultaneously cause the temperature
difference to disappear.  A host of pedagogical problems of this kind seem ideally suited
for analysis through the $\phi^4$  model and are recommended for further research.

\section {Acknowledgments}

We thank Carol Hoover ( Ruby Valley ) and Clint Sprott ( University of Wisconsin-Madison )
for useful discussions and help with the Figures. Puneet Patra ( Kharagpur ) provided several
useful comments on an earlier draft of this manuscript.  We wish to thank both Harald Posch
and Stefano Ruffo for their remarks and suggestions.  Their ideas
were useful in improving our description of the symmetry breaking found in the antisymmetric
solutions of the type illustrated in Figure 5.  Shortly after this paper was prepared K. A.
published a detailed account of the structure of the Lyapunov spectrum for the
$\phi^4$ model: ``Stable and Unstable Periodic Orbits in the One-Dimensional $\phi^4$ Theory'',
Physical Review E {\bf 94}, 042209 (2016).  His work
was supported in part by the Grant-in-Aid for Scientific Research  (\#15K05217) from the
Ministry of Education, Culture, Sports, Science, and Technology of Japan, as well as a
grant from Keio University.


\begin{thebibliography}{99}

\bibitem{b1}  K. Aoki and D. Kusnezov, ``Bulk Properties of Anharmonic Chains in Strong
              Thermal Gradients: Nonequilibrium $\phi^4$ Theory'', Physics Letters A
              {\bf 265}, 250-256 (2000).

\bibitem{b2}  K. Aoki and D. Kusnezov, ``Nonequilibrium Steady States and Transport
              In the Classical Lattice $\phi^4$ Theory'', Physics Letters B {\bf 477},
              348-354 (2000).

\bibitem{b3}  B. Hu, B. Li, and H. Zhao, ``Heat Conduction in One-Dimensional
              Nonintegrable Systems'', Physical Review E {\bf 61}, 3828-3831 (2000).

\bibitem{b4}  B. L. Holian, W. G. Hoover, and H. A. Posch, ``Resolution of Loschmidt's
              Paradox: The Origin of Irreversible Behavior in Reversible Atomistic
              Dynamics'', Physical Review Letters {\bf 59}, 10-13 (1987).

\bibitem{b5}  B. Moran, W. G. Hoover, and S. Bestiale, ``Diffusion in a Periodic Lorentz   
              Gas'', Journal of Statistical Physics {\bf 48}, 709-726 (1987).

\bibitem{b6}  Wm. G. Hoover, H. A. Posch, K. Aoki, and D. Kusnezov, ``Remarks on          
              NonHamiltonian Statistical Mechanics: Lyapunov Exponents and Phase-Space
              Dimensionality Loss'', Europhysics Letters {\bf 60}, 337-341 (2002).

\bibitem{b7}  W. G. Hoover and H. A. Posch, ``Direct Measurement of Lyapunov Exponents'',
              Physics Letters A {\bf 113}, 82-84 (1985).

\bibitem{b8}  W. G. Hoover and H. A. Posch, ``Direct Measurement of Equilibrium and
              Nonequilibrium Lyapunov Spectra'', Physics Letters A {\bf 123}, 227-230
              (1987).

\bibitem{b9}  K. Aoki and D. Kusnezov, ``Lyapunov Exponents and the Extensivity of           
              Dimensional Loss for Systems in Thermal Gradients'', Physical Review E
              {\bf 68}, 056204 (2003) .

\bibitem{b10} H. A. Posch and W. G. Hoover, ``Large-System Phase-Space Dimensionality
              Loss in Stationary Heat Flows'', Physica D {\bf 187}, 281-293 (2004).

\bibitem{b11} W. G. Hoover and C. G. Hoover, {\it Simulation and Control of Chaotic
              Nonequilibrium Systems}, Section 7.7, pages 204-205 (World Scientific
              Publishing, Singapore, 2015).

\bibitem{b12} F. Ricci-Tersenghi, ``The Solution to the Challenge in `Time-Reversible
              Random Number Generators' by Wm. G. Hoover and Carol G. Hoover'',
              arXiv.1305.1805.

\bibitem{b13} W. G. Hoover, B. L. Holian, and H. A. Posch, ``Comment I on `Possible
              Experiment to Check the Reality of a Nonequilibrium Temperature' '',
              Physical Review E {\bf 48}, 3196-3198 (1993).

\bibitem{b14} Wm. G. Hoover, K. Aoki, C. G. Hoover, and S. V. De Groot, ``Time-Reversible
              Deterministic Thermostats'', Physica D {\bf 187}, 253-267 (2004).

\end{thebibliography}
\end{document}